\documentclass[letterpaper, 10 pt, conference]{ieeeconf} 

\IEEEoverridecommandlockouts   

\usepackage{bmpsize}\usepackage[utf8]{inputenc}
\usepackage{graphicx}\usepackage[utf8]{inputenc}
\usepackage{amssymb,amsmath}

\makeatletter
\renewcommand*\env@matrix[1][\arraystretch]{%
	\edef\arraystretch{#1}%
	\hskip -\arraycolsep
	\let\@ifnextchar\new@ifnextchar
	\array{*\c@MaxMatrixCols c}}
\makeatother

\usepackage{float}
\usepackage{amssymb}
\usepackage{soul}
\graphicspath{{Images/}}
\usepackage[colorlinks={true},linkcolor={black},citecolor={black}, urlcolor={black}]{hyperref}
\emergencystretch\maxdimen
\newtheorem{assumption}{Assumption}
\newtheorem{theorem}{Theorem}
\newtheorem{proofoftheorem}{Proof of Theorem}

\title{\LARGE \bf 
 Feedback Regularization and Geometric PID Control for Robust Stabilization of a Planar Three-link Hybrid Bipedal Walking Model
 }

\author{W.M.L.T.Weerakoon$^{1}$, T.\,W.\,U.\,Madhushani$^{2}$, D.H.S.Maithripala$^{3}$ and J.M.Berg$^{4}$
	\thanks{$^{1}$Department of Mechanical Engineering, University of Peradeniya, KY 20400, Sri Lanka.
			{\tt\small Lasithaweera@eng.pdn.ac.lk}}%
	\thanks{$^{2}$Postgraduate and Research Unit, Sri Lanka Technological Campus, CO 10500, Sri Lanka.
			{\ttfamily\small udarim@sltc.lk}}%
	\thanks{$^{3}$Department of Mechanical Engineering, University of Peradeniya, KY 20400, Sri Lanka.
		{\tt\small smaithri@pdn.ac.lk}}%
	\thanks{$^{4}$Department of Mechanical Engineering, Texas Tech University, TX 79409, USA.
		{\tt\small jordan.berg@ttu.edu}}%
}

\begin{document}
	\maketitle
	\begin{abstract}
	This paper applies a recently developed geometric PID controller to stabilize a  three-link planar bipedal hybrid dynamic walking model. The three links represent the robot torso and two kneeless legs, with an independent control torque available at each hip joint. The geometric PID controller is derived for fully actuated mechanical systems, however in the swing phase the three-link biped robot has three degrees of freedom and only two controls. Following the bipedal walking literature, underactuation is addressed by choosing two ``virtual constraints'' to enforce, and verifying the stability of the resulting two-dimensional zero dynamics. The resulting controlled dynamics do not have the structure of a mechanical system, however this structure is restored using ``feedback regularization,'' following which geometric PID control is used to provide robust asymptotic regulation of the virtual constraints. The proposed method can tolerate significantly greater variations in inclination, showing the value of the geometric methods, and the benefit of integral action.
\end{abstract}
\allowdisplaybreaks


%
%
%
%
%


\section{Introduction}\label{sec:Intro}

\begin{figure}[b!]
	\centering
	\begin{tabular}{c}
		 \includegraphics[width=2.8in]{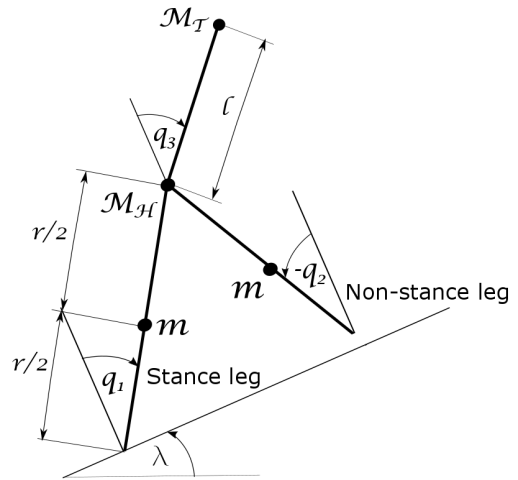}
	\end{tabular}
		\caption{Planar three link biped walker on an inclined plane with lumped masses. The robot consists of a torso and two equal length legs connected to the torso at the hip. A control moment is applied at the hip joint between each leg and the torso.}
\label{Fig:model}
\end{figure}

Insight into bipedal walking may be gained through the study of the simple three-link planar model shown in Fig. \ref{Fig:model}, consisting of a torso and two straight, unjointed, legs \cite{McGeer, mcgeer_passivedynamics, grizzlebook}. Each leg is connected to the torso at a hip joint, and each leg is actuated by an independent hip moment. The free end of each leg is referred to as a foot. A walking gait alternates between a \textit{stance phase}, in which both feet are on the ground, and a \textit{swing phase}, in which the foot of the \textit{support leg} is in contact with the ground, and the foot of the \textit{swing leg} is not in contact with the ground. The stance phase is assumed to be instantaneous, and is associated with impulsive momentum transfer between the links as the swing leg impacts the ground and becomes the support leg \cite{grizzlebook, grizzle_impact}. The impact dynamics associated with the stance phase are discrete, while the rigid-body dynamics associated with the swing phase are continuous. Thus the complete description combining both phases is a hybrid model. This model requires certain non-physical assumptions, such as neglect of ``scuffing'' -- that is, incidental contact between the swing foot and the ground during the swing phase. Other assumptions include that the support foot may rotate freely, but cannot slip.

The three-link hybrid model is used in \cite{grizzle_impact} to analyze the stability of planar walking gaits under a certain class of control laws. In this paper the approach of \cite{grizzle_impact} is used to analyze the stability of an alternate class of control laws. In \cite{grizzle_impact}, the desired walking gait is parametrized using the stance leg angle $q_1$, which is assumed to increase monotonically in time. That is, the desired swing leg angle is written as $q_{2,d} = \eta_2(q_1)$ and the desired torso angle is written as $q_{3,d} = \eta_3(q_1)$. Then an output function is defined as $y(q) = [q_2 - \eta_2(q_1), q_3 - \eta_3(q_1)]^T$, and the equation $y(q) = [0,0]^T$ defines the output zeroing manifold. Finally, a switching hypersurface is defined by $S(q) = 0$. A fundamental assumption is that system trajectories intersect the switching surface, $\mathcal{S}$, transversely. When the trajectory crosses the switching surface, the swing leg is assumed to have contacted the ground, and the old stance leg becomes the new swing leg. A discrete-time stability analysis can be carried out using the switching surface as a Poincare section. In \cite{grizzle_impact}, partial-feedback linearization (PFL) followed by a nonlinear finite-time feedback law is used to guarantee finite-time convergence of the output function to zero. That is, the system trajectories converge to the output zeroing manifold in finite time, and remain on that manifold for the remainder of the swing phase. The convergence time is chosen so that the system trajectory converges to the output zeroing manifold before crossing the switching surface, so that the Poincare section can be taken as the intersection of the switching surface and the output zeroing manifold. This significantly simplifies their stability analysis. 

Different methods incorporating feedback linearization have been used extensively to control biped robots on a flat perfectly horizontal ground \cite{grizzlebook, grizzle_impact, grizzleperiodic, grizzle_HZD_planar, Choi, Ames2017, threelinkTraj} as well as on rough terrain \cite{L2Gain_unknownTerrain_Tedrake, Grizzle_DiscreteTerrain_Walking, Stepping_stone}. Decentralized control methods have been used in \cite{Decentralized_akbari} for stabilization of periodic orbits in walking gaits. Human inspired techniques as well as bionic methods also have been experimented in \cite{raibert1986legged, Yadukumar2013, AmesHuman_insp, Zhou_bio_insp} to achieve stable periodic gaits.

For the current paper, feedback regularization and geometric PID control take the place of partial-feedback linearization and finite-time control. This approach to exponential tracking for underactuated mechanical systems was introduced in \cite{udariHoop,udarispherical} and a feedback regularization based geometric PD control for a class planar of three link was tested in \cite{Lasitha_biped}. Geometric PID control as presented in \cite{maithripala_geometric, MaithripalaPID, Gorez_PID} provides a powerful and intuitive robust control design method for fully actuated mechanical systems. To apply this method to underactuated systems, the configuration variables are replaced by a number of output functions equal to the number of controls. This takes care of the underactuation, however the coordinate change typically destroys the mechanical system structure. Then feedback is used to recover the form of a mechanical system -- this is the process of \textit{feedback regularization}. Now a geometric PID control may be designed. The resulting closed-loop system is stable if the zero dynamics resulting from the output functions are stable. 

Use of feedback regularization plus geometric PID guarantees \textit{asymptotic} convergence of trajectories to the output zeroing manifold, but the trajectories do not actually reach the manifold in finite time. However we show that given any small neighborhood, $\mathcal{Z}_\delta$, of the zerodynamic manifold, $\mathcal{Z}$, there exists PID controller gains that guarantee that if you start in $\mathcal{Z}_\delta$ then you will repeatedly return to this neighborhood. The underlying motivation for using this controller is that the robustness provided by the PID control -- particularly the integral action -- will result in a larger region of attraction for the walking gait. Note that while the planar three-link biped model is simple, it provides a valuable test bed to demonstrate the robust properties of the proposed control framework.

Section \ref{sec:Theory} presents the hybrid dynamical model used in \cite{grizzle_impact} for the three-link planar biped model shown in Figure \ref{Fig:model}. For convenience, and validity of comparison, we use the same output function as \cite{grizzle_impact}. Section \ref{sec:controller} derives the feedback regularization and geometric PID controller for this model, and analyzes the stability of the resulting gaits. Section \ref{sec:simulations} present simulation results for the proposed controller on planes with various unknown, but constant, inclination, and in the presence of substantial parameter variations showing that the geometric control approach, and that the use of integral action provides an important degree of robustness.  We consider both uphill and downhill motion.

\section{Biped Robot Dynamics}\label{sec:Theory}

In this section we briefly discuss the mathematical  model of the planar three-link biped robot on an inclined plane of unknown inclination $\lambda$. The model was originally introduced in \cite{grizzle_impact} and serves as the initial test bed for control algorithms for a class of biped robots that are characterized by dynamic stability \cite{grizzle_impact}. As shown in Fig \ref{Fig:model} the masses of the links are lumped. The mass of each leg is denoted by $m$, the mass of the torso at the hip joint end is denoted by $M_H$ and the mass of the torso at the other end is denoted by $M_T$. The coordinates $q\triangleq (q_1,q_2,q_3)$ as indicated in Fig \ref{Fig:model} are used to prescribe the configuration of the robot and we refer the reader to \cite{grizzle_impact} for a detailed analogous derivation of the mathematical model for a three-link biped robot on a flat surface.

The motor torques $u=[u_1,u_2]^T$ will be applied between the torso and the two legs; $u_1$ and $u_2$ are the torques associated with the stance leg and the non stance leg respectively. Typically, the dynamics of a biped robot fall into two distinct phases, the single stance (swing phase) and the double stance (impact phase) and thus the complete dynamic behavior of the biped is a hybrid of these two phases \cite{grizzlebook, grizzle_impact}. 

During the swing pahse, one of the robot's legs is implanted on the ground and the other leg is swinging. The dynamics of the system during this phase is described by the Euler Lagrange equations,
\begin{align}
\mathbb{I}\nabla_{\dot{q}}\dot{q}&= G + B\:u \label{eq:DynaSS}
\end{align}
where $\nabla_{\dot{q}}\dot{q}$ is the Levi-civita connection associated with the mass-inertia tensor $\mathbb{I}$ that is given by,
{\small
\begin{align*}
\mathbb{I} (q) = 
\begin{bmatrix} [2]
\dfrac{\begin{smallmatrix} 4M_{H}r^2 + 4M_{T}r^2\\ + 5mr^2 \end{smallmatrix}}{\begin{smallmatrix} 4 \end{smallmatrix}} 
 &   
-\dfrac{\begin{smallmatrix} mr^2\cos(q_{1} - q_{2})\end{smallmatrix}}{\begin{smallmatrix} 2 \end{smallmatrix}}&    
\begin{smallmatrix} M_{T}lr\cos(q_{1} - q_{3}) \end{smallmatrix}\\
-\dfrac{\begin{smallmatrix} mr^2\cos(q_{1} - q_{2})\end{smallmatrix}}{\begin{smallmatrix} 2 \end{smallmatrix}}&
\dfrac{\begin{smallmatrix} mr^2 \end{smallmatrix}}{\begin{smallmatrix} 4 \end{smallmatrix}} & 
\begin{smallmatrix} 0 \end{smallmatrix}\\
\begin{smallmatrix} M_{T}lr\cos(q_{1} - q_{3}) \end{smallmatrix} &
\begin{smallmatrix}0 \end{smallmatrix}&
\begin{smallmatrix} M_{T}l^2
\end{smallmatrix}
\end{bmatrix}
\end{align*}
}
the gravitation interactions,
\begin{align*}
G=
  \begin{bmatrix}[1.4]
  \frac{gr(2M_{H} + 2M_{T} + 3m)\sin(q_{1} - \lambda)}{2} \\
 -\frac{gmr\sin(q_{2} - \lambda)}{2} \\
 \begin{smallmatrix} M_{T}gl\sin(q_{3} - \lambda) \end{smallmatrix}\\
   \end{bmatrix}
\end{align*}\\
and input matrix,
{\small
\begin{align*}
B=
  \begin{bmatrix}[1.2]
 -1 &  0\\
  0 & -1\\
  1 &  1\\
   \end{bmatrix}. \label{eq:Bmatrix}
\end{align*}
}

The explicit expression for $\mathbb{I}\nabla_{\dot{q}}\dot{q}$ in terms of the coordinates $q\triangleq (q_1,q_2,q_3)^T$ is provided in equation-(\ref{eq:Levi-DynaSS}) of the appendix for easy reference. We also refer the reader to the excellent texts \cite{Riemannian_bullo, SingleRiemannean_stoica, godinho2012_riemannian} for supplementary material on mechanical systems on Riemannian manifolds.

Letting $x\triangleq (q,\dot{q})\in \mathcal{X}$ we will denote the state space representation of the swing phase as,
\begin{align}
\dot{x}&=
f(x)+g(x) \: u,
\end{align}
where $f(x)$ and $g(x)$ are found using (\ref{eq:DynaSS}). 

The impact phase is the phase during the infinitesimally small time when both the legs will be on the ground. It happens when the swing leg comes and strikes the ground with the assumption of no slip and no rebound conditions. At the end of the double stance case the robot will move on to the single stance again and this periodic cycle is known as a \textit{walking gait}. Therefore at the end of the double stance phase, the labelling of the two legs must interchange, and hence one must switch the legs, relabelling the stance leg as the swing leg and the swing leg as the stance leg. A complete derivation of the impact model is found in \cite{grizzle_impact} and it is used to derive the reset map for the new step. Taking $x^{-}$ as the instance just prior to the impact and $x^{+}$ as the instance just after the impact, the reset map \cite{grizzle_impact} can be stated as,
\begin{align}
x^+ &= \Delta (x^{-})
\end{align}
An overview of the reset map is provided in section-\ref{Secn:ResetMap} of the the Appendix for completeness.

The walking gait is a combination of the above mentioned swing phase and the impact phase. This combination is referred to as the \textit{hybrid model of walking} and can be stated as,
\begin{align}
\sum & := \Bigg\{\begin{matrix}[1.8]
\dot{x} = f(x)+g(x) \: u \:\: &; x^-\notin\mathcal{S}\\
x^+ = \Delta (x^{-})  &; x^-\in\mathcal{S} 
\end{matrix}
\label{eq:hybridsystem}
\end{align}
where $\mathcal{S}$ is the switching surface defined as,
\begin{align}
\mathcal{S} := \{(q,\dot{q}) \in \mathcal{X} \: \: \: | \: \: \: q_1=q_{1_{ref}} \}, \label{eq:switchingsurf}
\end{align}
with $q_{1_{ref}}$ being a preset constant angle for this work.
When the tracking error is identically equal to zero,  
the switching surface (\ref{eq:switchingsurf}) indicates the instance when the swing leg reaches the ground. If the tracking error is not zero then when the trajectories cross $\mathcal{S}$ then either the swing leg may not have reached the ground or leg scuffing may occur as $0<|q_2+q_1|\leq \delta $ for some $\delta>0$. Specifically we assume the following;
\begin{assumption}
	When $\delta$ is sufficiently small there exists a mechanism to initiate contact with the walking surface or avoid leg scuffing without affecting the impact map. 
\end{assumption}

Thus in effect assuming that it is enough to reach sufficiently close to the zerodynamics manifold in order to complete the step.

\section{Stabilization of a Periodic Gait} \label{sec:controller}
%

The simplest way to idealize walking is to achieve posture control and swing leg advancement. It is shown in \cite{grizzle_impact, grizzle_HZD_planar} that an input-output linearization based finite time stabilizing controller is sufficient  to achieve an asymptotically stable periodic gait for the class of bipeds used in this work. Following \cite{grizzle_impact} we select the output function for our planar three-link biped robot to be, $ q_e = ( q_1 + q_2,q_3 - q_{3_{ref}})^T \in G_e=\mathbb{S} \times \mathbb{S} $ where $q_{3_{ref}}$ is a constant. Notice that the state space $\mathcal{X}$ is now diffeomorphic to $TG_e\times T\mathbb{S}$ with local coordinates $x=(q_e,\dot{q}_e,q_1,\dot{q}_1)$. The zerodynamics of the system then evolve on  $\mathcal{Z}\triangleq \{(0,0)\}\times TS\equiv TS$. 

In this section we will develop a nonlinear controller that will  exponentially stabilize the output during the swing phase and show that this is sufficient to robustly achieve an asymptotically stable periodic gait under the Assumption 1.
The controller is based on the notion of feedback regularization control plus intrinsic PID control introduced in \cite{udariHoop,udarispherical, MaithripalaPID}.

From (\ref{eq:DynaSS}) we see that the dynamic model of the biped robot can be written down in the coupled form:  
\begin{align}
\mathbb{I}_e  \dot{\omega}_e + \tau_e(\omega_e,\omega_1) + \tau_g^{e} &= \tau_u^e \label{eq:outputEQ}
\\
\mathbb{I}_z  \dot{\omega}_1 +\tau_z(\omega_1,\omega_e) + \tau_g^{z} &=   \tau_u^z
\label{eq:actuatorEQ}
\end{align} 
where  $\dot{q}_1=\omega_1$ and,\\
\begin{align*}
\mathbb{I}_{e} (q_s)&=	
\begin{bmatrix} [1.5]
\begin{smallmatrix}
l^2(4\, {M_H} + 2\, {M_T}(1- \cos\!\alpha)\\
+m(3- 2\cos\beta)\:) \end{smallmatrix}& \begin{smallmatrix}
0 \end{smallmatrix}\\
\begin{smallmatrix}
0 \end{smallmatrix}& \begin{smallmatrix}
r^2(4\, {M_H} + 2\, {M_T}(1- \cos\!\alpha)\\
+m(3- 2\cos\beta)\:)\end{smallmatrix}
\end{bmatrix} 
\\
\mathbb{I}_z (q_s) &=
\frac{r^2}{4}
\begin{smallmatrix}
{\left(4\, {M_H} + 2\, {M_T} + 3\, m - 2\, {M_T}\, \cos\!\left(\alpha\right) - 2\, m\, \cos\!\left(\beta\right)\right)}
\end{smallmatrix}
\end{align*}
with $\alpha \triangleq 2(q_1-q_{3_{ref}}-q_{e_1})$, $\beta \triangleq 2(2q_1-q_{e_2}) $ and  $q_s\equiv (\alpha,\beta)$, and the control inputs,
\begin{align*}
\tau_u^e=&
B_e u
\\
\tau_u^z=&
\begin{pmatrix}
	\begin{smallmatrix}
		- \frac{r}{l} \cos(\frac{\alpha}{2}) - 1 \end{smallmatrix} & \begin{smallmatrix} - 2 \cos(\frac{\beta}{2}) - \frac{r}{l} \cos(\frac{\alpha}{2}) \end{smallmatrix}
\end{pmatrix}
u
\end{align*}
where,
\begin{align*}
{B_e}_{11} =& 
\begin{smallmatrix}
\frac{4 M_H + 3 m - 2 m \cos({\beta})}{M_T} + \frac{4\big( r +  l \cos(\frac{\alpha}{2})\big)}{r} \end{smallmatrix} \\
{B_e}_{12} =& \begin{smallmatrix} \frac{4 M_H + 3 m - 2 m \cos({\beta})}{M_T} + \frac{4\big( r +  l \cos(\frac{\alpha-\beta}{2}) + l \cos(\frac{\alpha+\beta}{2})\big)}{r}
\end{smallmatrix}\\
{B_e}_{21} =&\begin{smallmatrix}
- \frac{4 \big(l + r \cos(\frac{\alpha}{2})\big) \big(2 \cos(\frac{\beta}{2}) + 1\big)}{l}\end{smallmatrix} \\
{B_e}_{22} =& \begin{smallmatrix} - \frac{4\big(4 M_H + 2 M_T + 5 m - 2 M_T \cos({\alpha}) + 2 m \cos(\frac{\beta}{2})\big)}{m} \\
- \frac{4\big( r \cos(\frac{\alpha}{2}) +  r \cos(\frac{\alpha-\beta}{2}) +  r \cos\frac{\alpha+\beta}{2}\big)}{l} \end{smallmatrix}
\end{align*}

Let $\omega_s\triangleq \dot{q}_s=(\dot{\alpha},\dot{\beta})$. The quadratic velocity dependent forces $\tau_e(\omega_e,\omega_1), \:\tau_z(\omega_1,\omega_e) $ and the gravitational interaction terms ${\tau_g}^{e},\:{\tau_g}^{z}$ are provided in equations (\ref{eq:tau_e})--(\ref{eq:tau_g1}) in the Appendix.
Here (\ref{eq:outputEQ}) represents the error dynamics of the system and  when the output $q_e(t)$ is restricted to zero (\ref{eq:actuatorEQ}) represents the zerodynamics of the system.

Following the feedback regularization approach proposed in \cite{udariHoop, udarispherical} we choose the regularizing plus potential shaping controls,
\begin{align}
\tau_u^e = \tau_g^e -\mathbb{I}_e^{-1}\Gamma(\omega_s)\omega_e + \tilde{\tau},
\end{align}
where
\begin{align*}
\Gamma (q_s,\omega_s)=& 
\mathbb{I}_e^{-1}\begin{bmatrix} [2.2]
\begin{smallmatrix} M_T l^2 \sin (\alpha) \dot{\alpha} \\
+  m l^2 \sin (\beta) \dot{\beta} \end{smallmatrix} & 
\begin{smallmatrix} m l^2 \sin(\beta)\dot{\alpha} \\
- M_T r^2 \sin(\alpha) \dot{\beta}   \end{smallmatrix} \\
\begin{smallmatrix}  -m l^2 \sin(\beta)\dot{\alpha} \\
 + M_T r^2 \sin(\alpha) \dot{\beta}  \end{smallmatrix} & 
 \begin{smallmatrix} M_T r^2 \sin (\alpha) \dot{\alpha} \\
 +  m r^2 \sin (\beta) \dot{\beta}\end{smallmatrix} 
\end{bmatrix}
\end{align*}
to give the error dynamics (\ref{eq:outputEQ}) the structure of a simple mechanical system on the Lie group 
$G_e=\mathbb{S}\times \mathbb{S}$:
\begin{align}
{\mathbb{I}_{e}}(q_s) {\nabla}^e_{\omega_s}{\omega_e} = & \tilde{\tau}_u + \tau_d
\label{eq:recoveredMechStructure_system}
\end{align} 
where ${\nabla}^e_{\omega_s}{\omega_e}$ is the unique Levi-Civita connection associated with the inertia tensor $\mathbb{I}_e$ that is explicitly given by,
\begin{align}
{\nabla}^e_{\omega_s}{\omega_e}&=\dot{\omega}_e+\Gamma(\omega_s)\omega_e.
\end{align}
The term $\tau_d$ is introduced here to represent any  moments due to unmodelled disturbances, parameter uncertainties, and the effects due to the lack of accurate information of the inclination plane.
This structure allows us to use the intrinsic PID controller proposed in \cite{MaithripalaPID} to ensure that the error dynamics converge to zero exponentially, provided that the zerodynamics of the system remains bounded. This PID controller takes the form,
\begin{align}
&\tilde{\tau}_u =  -\mathbb{I}_s(k_p \eta_e + k_d \omega_e + k_I \omega_I)
\label{eq:PID}
\\
&{\mathbb{I}_{e}}(q_s) {\nabla}^e_{\omega_s}{\omega_I} =  \mathbb{I}_e \eta_e
\label{eq:integralTerm}
\end{align} 
where, 
\begin{align}
{\nabla}^e_{\omega_s}{\omega_I}&=\dot{\omega}_I+\Gamma(\omega_s)\omega_I,\\
\mathbb{I}_e \eta_e &= 
\begin{bmatrix} 
\sin{\left(q_3 - {q_3}_{ref}\right)}\\
\sin({q_1+q_2})
\end{bmatrix}. \label{eq:v_fun}
\end{align}
Notice that the controller does not require the knowledge of the inclination of the plane. However, it can be shown that the system parameters should satisfy, $\frac{\sin(q_{3_{ref}}-\lambda)}{\sin(\lambda)}\geq \frac{(M_T + M_H + m)r}{M_T l}$ for static stability. Hence, it should be noted that the selection of the desired torso angle $q_{3_{ref}}$ is not arbitrary for a particular range of inclinations. 

Considering the unique Levi-Civita connection of $\mathbb{I}_z(\theta_z)$ that is explicitly given by 
\begin{align*}
\mathbb{I}_z(\theta_1)\nabla^z_{{\omega_1}}{\omega_1}= \mathbb{I}_z(\theta_1) \dot{\omega}_1 - M_Trl\sin(\theta_1)\omega_1^2, 
\end{align*}
we see that the dynamics (\ref{eq:actuatorEQ}) also have the mechanical system structure:
\begin{align}
\mathbb{I}_z(\theta_1)\nabla^z_{{\omega_1}}{\omega_1} = -\tau_g^{z}-\tau_z(\omega_e,\omega_1).\label{eq:zero_dynamicsMech}
\end{align}
The gravitational interaction term $\tau_g^{z}$ and the quadratic velocity term $\tau_z(\omega_e,\omega_1)$ satisfy the conditions specified in Assumption 1 of \cite{udariHoop} and hence from Theorem 1 of \cite{udariHoop} we have that
during the swing phase for any initial condition set $\mathcal{X}_e^0\times \mathcal{Z}^0\subset TG_e\times \mathcal{Z}$ there exists PID controller gains for the controller (\ref{eq:PID}) -- (\ref{eq:integralTerm}) such that $\lim_{t\to \infty }(\eta_e(t),\dot{q}_e(t))=(0,0)\in TG_e$ exponentially while ensuring that $|\dot{q}_1(t)|$ remains bounded even in the presence of bounded parameter uncertainty. 
Using this result we prove following theorem in the appendix. \\
\begin{theorem}\label{Theom:Main}
Given any small $\delta>0$ there exists PID controller gains for the controller (\ref{eq:PID}) -- (\ref{eq:integralTerm}) such that
the trajectories of the closed loop system (\ref{eq:DynaSS}) satisfies $x(t^-_k)\in \mathcal{Z}_\delta$, where
\begin{align}
\mathcal{Z}_\delta =\{(q_e,\dot{q_e})\in \mathcal{X} | \sqrt{||\eta_e||^2+||\dot{q}_e||^2}\leq \delta\},
\end{align}
for some increasing sequence of  time steps $t^-_0<t^-_1<\cdots<t^-_k<\cdots$. 
\end{theorem}

We point out that $\mathcal{Z}\subset \mathcal{Z}_\delta$ and that $\mathcal{Z}_\delta$ is a small $\delta$-neighborhood of the zerodynamics manifold $\mathcal{Z}$.
Thus what this theorem implies is that if the biped robot starts with an initial condition that is close to $\mathcal{Z}$ then the trajectories of the closed-loop system visit this neighborhood during every step of the robot. In the following we will show numerically that these repeated trajectories converge asymptotically to a periodic orbit. Thus ensuring the asymptotic stabilization of a walking gait.


\section{Simulation Results} \label{sec:simulations}
In this section we present the simulation results for the proposed feedback regularization and geometric PID controller for the planar three-link biped robot walking on an inclined plane. The nominal parameters of the robot used in this work are $m=1kg$, $M_T=3kg$, $M_H= 1kg$, $l=0.75m$ and $r=1m$ which assumes a planar three link biped carrying a substantial load on its torso. The maximum inclination of the plane for equilibrium for these parameters is $\lambda_{max}=26.7^{\circ}$. Thus for the simulations we use an inclination of $\lambda=25^{\circ}$. The switching surface for all the simulations were set at $q_{1_{ref}} = 15 ^{\circ}$ and for all the simulations we have picked $\delta=1.5\times 10^{-3}$. The controller gains we use are $K_p = 1500$, $K_d=1250$ and $K_I = 120$.

\begin{figure}[h!]
	\centering
	\begin{tabular}{c}
		\includegraphics[width=3.1in]{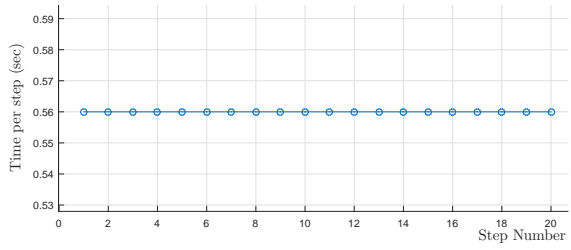}\\
		{\footnotesize a) Step time}\\
	    \includegraphics[width=3.0in]{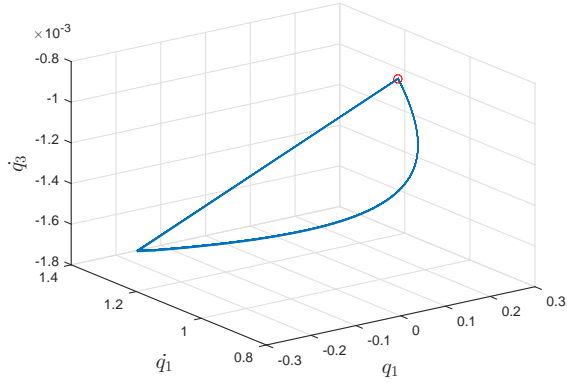}\\
		{\footnotesize b) Trajectory projected onto $(q_1,\dot{q_1},\dot{q_3})$}\\
	\end{tabular}
		\caption{Simulation of 20 steps for initial conditions that begin in $\mathcal{Z}_\delta$ with $\dot{q}_1=1.168 \mathrm{rad/s}$. This shows that the periodic orbit that gives rise from $\mathcal{Z}_\delta$ converges asymptotically to a closed orbit and has a stable periodic gait with a step time of 0.56 s. The circle in b) indicates the initial point.}
	\label{Fig:stepTime}
\end{figure}

\begin{figure}[h!]
	\centering
	\begin{tabular}{c}
		\includegraphics[width=3.4in]{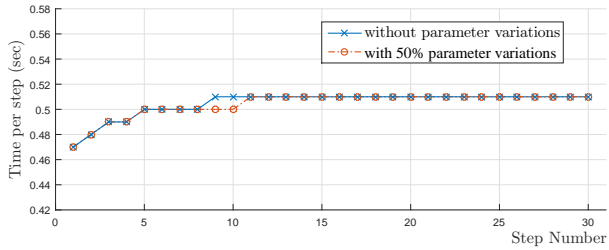}\\
		{\footnotesize a) Step time}\\
		\includegraphics[width=3.4in]{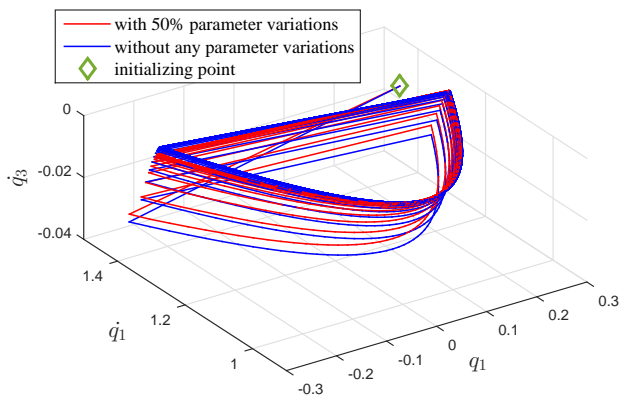}\\
		{\footnotesize b) Trajectory projected onto $(q_1,\dot{q_1},\dot{q_3})$}
	\end{tabular}
	\caption{Simulation of 30 steps with and without parameter variations climbing up a plane with an inclination of $\lambda=25^{\circ}$ initialized with a pre-impact $x^-(0)=(11.25^{\circ},-15^{\circ},112^{\circ},1.3\, \mathrm{rad/s},-1.4\,\mathrm{rad/s},0)$ and the desired torso angle set to $110^{\circ}$.}
	\label{Fig:robustness}
\end{figure}

\begin{figure}[h!]
	\centering
	\begin{tabular}{c}
			\includegraphics[width=3.1in]{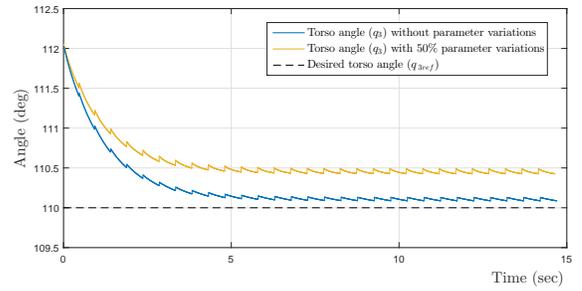}\\
		{\footnotesize a) Variation of torso angle with time} \\
		\includegraphics[width=3.0in]{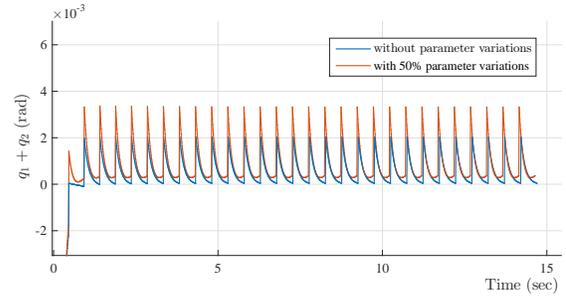}\\
		{\footnotesize b) Variation of ($q_1+q_2$) with time} 
	\end{tabular}
	\caption{Simulation of 30 steps which is initialized with pre-impact $x^-(0)=(11.25^{\circ},-15^{\circ},112^{\circ},1.3\, \mathrm{rad/s},-1.4\,\mathrm{rad/s},0)$. The desired torso angle was set to $110^{\circ}$. This illustrates the robustness of the controller to the parameter variations and it is seen that the output will visit $\mathcal{Z}_\delta$ repeatedly; hence asymptotically stable periodic gait. }
	\label{Fig:outputTracking}
\end{figure}

The existence of a periodic orbit, with the trajectory projected on $(q_1,\dot{q_{1}}, \dot{q}_3)$, and the step times are shown in Fig. \ref{Fig:stepTime} which clearly illustrates the convergence of the trajectory to a periodic orbit. The desired torso angle was set at $q_{3_{ref}}=105^{\circ}$ for this simulation. We also show the robustness of the periodic orbit to parameter uncertainties as large as 50\% in Fig. \ref{Fig:robustness}. The output error tracking for the torso and leg angles for this case are shown in Fig. \ref{Fig:outputTracking} and it clearly shows that there exists a periodic orbit that reaches $\mathcal{Z}_\delta$ at the end of each step.

\begin{figure}[h!]
	\centering
	\begin{tabular}{c}
		\includegraphics[width=3.3in]{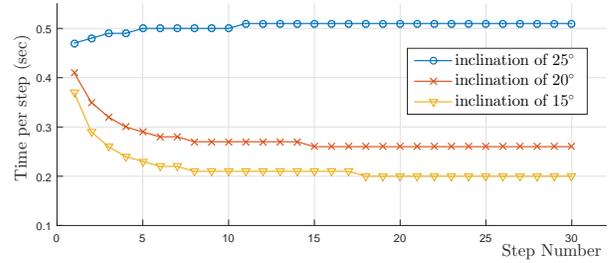}\\
		{\footnotesize a) Step time} \\
		\includegraphics[width=3.0in]{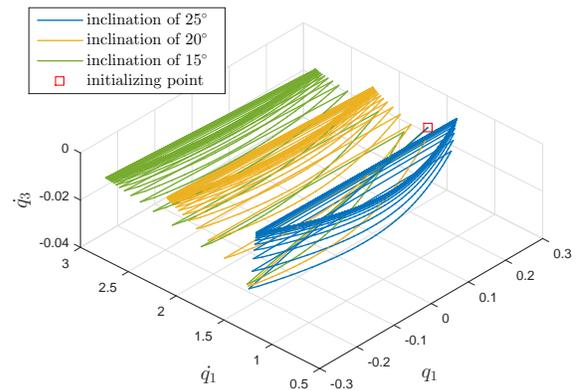}\\
		{\footnotesize b) Trajectory projected onto $(q_1,\dot{q_1},\dot{q_3})$} 
	\end{tabular}
	\caption{Simulation for 30 steps for different inclinations of the plane, in the presence of parameter uncertainties as large as 50\%, initialized with pre-impact $x^-(0)=(11.25^{\circ},-15^{\circ},112^{\circ},1.3\, \mathrm{rad/s},-1.4\,\mathrm{rad/s},0)$. The controller was set assuming the inclination $\lambda=25^{\circ}$ and desired torso angle was set to $110^{\circ}$.}
	\label{Fig:variation_of_inclination}
\end{figure}

\begin{figure}[h!]
	\centering
	\begin{tabular}{c}
		\includegraphics[width=2.9in]{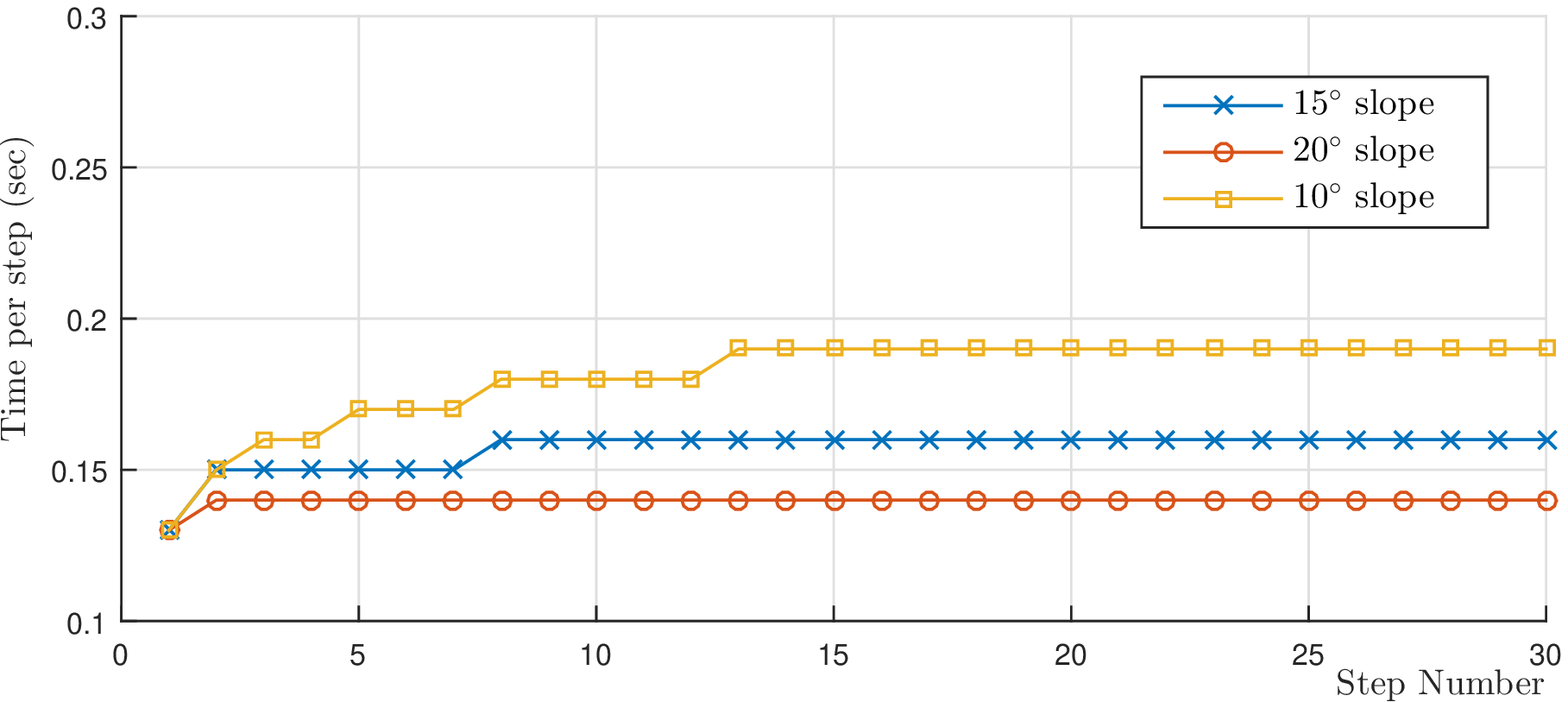}\\
		{\footnotesize a) Step time} \\
		\includegraphics[width=3.0in]{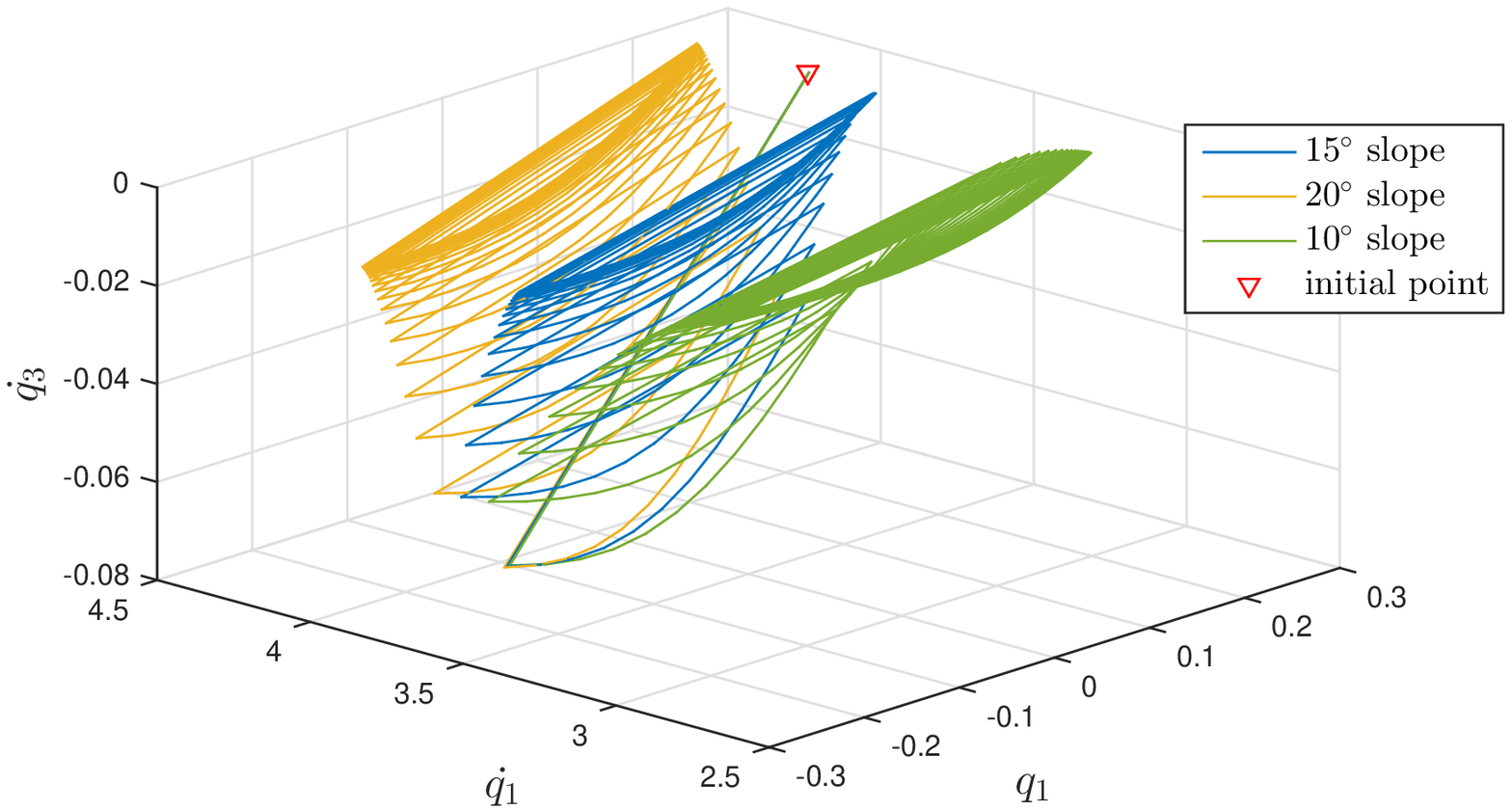}\\
		{\footnotesize b) Trajectory projected onto $(q_1,\dot{q_1},\dot{q_3})$} 
	\end{tabular}
	\caption{Simulation for 30 steps for different downhill slopes of the plane, in the presence of parameter uncertainties as large as 50\%, initialized with pre-impact  $x^-(0)=(12.85^{\circ},-11.25^{\circ},2^{\circ},4\, \mathrm{rad/s},-3.9\,\mathrm{rad/s},0)$. The controller was set assuming the inclination $\lambda=-15^{\circ}$ ($15^{\circ}$ slope) and desired torso angle was set to $0^{\circ}$.}
	\label{Fig:downhill}
	\end{figure}

We re-iterate that the accurate information of the inclination angle is not required in the controller and we demonstrate it in Fig.  \ref{Fig:variation_of_inclination} for uphill climbing and  Fig. \ref{Fig:downhill} for downhill descent. The simulation shows that even if the actual inclined angle is different from the angle set in the controller the asymptotically stable periodic orbit still exists even in the presence of large parameter uncertainties. The simulation also shows that the controller is applicable for uphill as well as downhill walking.

\section{Conclusion}
In this paper we propose a novel control strategy to obtain an asymptotically stable periodic gait for a class of under actuated planar biped robots. We have used the intrinsic notion of feedback regularization plus geometric PID control to ensure the robust existence of an asymptotically stable periodic orbit that can be made to approach the output zerodynamic manifold  arbitrarily closely by picking sufficiently large gains.  In particular it allows the robot to move up or down and inclined plane of unknown inclination. The controller does not require finite time stability. Since the notions developed are geometric and does not depend on coordinates it is expected that these ideas will equally be valid in the case of a more realistic five-link or seven-linked biped robot model and remains as future work.

\begin{appendix}
\subsection{Levi-Civita Connection of the three-linked biped robot}
The unique Levi-Civita connection corresponding to the inertia tensor $\mathbb{I}$ that corresponds to the Euler-Lagrange dynamics during the swing phase is explicitly computed to be,
{\small
\begin{align}
\mathbb{I}\nabla_{\dot{q}}\dot{q}&= \mathbb{I} \ddot{q} + \Gamma(q,\dot{q})\dot{q} 
\label{eq:Levi-DynaSS}
\end{align}}
where the non zero elements of $\Gamma(q,\dot{q})$ the connection matrix are given by,
{\small
\begin{align*}
\Gamma(q,\dot{q})_{12}&= -\dfrac{mr^2\sin(q_{1} - q_{2})\dot{q_{2}}}{2} \\
\Gamma(q,\dot{q})_{13}&= M_{T}lr\sin(q_{1} - q_{3})\dot{q_{3}}\\
\Gamma(q,\dot{q})_{21}&= \dfrac{mr^2\sin(q_{1} - q_{2})\dot{q_{1}}}{2}   \\
\Gamma(q,\dot{q})_{31}&= -M_{T}lr\sin(q_{1} - q_{3})\dot{q_{1}}.\\
\end{align*}}
\subsection{The reset map}\label{Secn:ResetMap}
The dynamic equation for the double stance model \cite{grizzle_impact} is as follows;
{\small
\begin{align}
D_{ds}(q_{ds})\ddot{q_{ds}} + C_{ds}(q_{ds},\dot{q_{ds}})\dot{q_{ds}} + G_{ds} &= B_{ds}u + \delta F_{ext} 
\end{align}
}where the explicit expressions for $D_{ds}(q_{ds})$, $C_{ds}(q_{ds},\dot{q_{ds}})$, $G_{ds}$ and $ B_{ds}$ can be found in the Appendix of \cite{grizzle_impact}.

The coordinate system for this case is taken to be $q_{ds}=[q;p_{h}]$, where  $p_{h}=[p_{hip}^{x};p_{hip}^{y}]$ are the Cartesian coordinates of the hip. Here the $\delta F_{ext}$ represents the impact force due to the striking of the non-stance leg on the ground. The impulsive contact force can be found as, $F_{ext} = \int_{t^-}^{t^+} \delta F_{ext} dt $.

At the instance just before the foot strike, let $\dot{q_{ds}}{^{-}}$ be the velocities and it is obtained from the single stance model. At this instance, the hip position $p_{h}^{-}= \gamma_{ds}(q )$ can be found using the body angle coordinates. The reaction force is at the end of the non-stance leg $p_{2}(q_{ds})$. The impact map,
\begin{align}
x^+ &= \Delta (x^{-})
\end{align}
where, {\small
\begin{align*}
\Delta (x^{-}) &=
\begin{bmatrix} [1.8]
\Delta_{q } q ^{-}\\ 
\Delta_{\dot{q }} (q^{-})\dot{q }{^{-}}
\end{bmatrix}\\
\Delta_{q }&=R\\
\Delta_{\dot{ }} (q ^{-}) &=
\begin{bmatrix} [1.5]
R & 0_{3\times2}
\end{bmatrix} 
\bar{\Delta}_{\dot{{q_{ds}}}}(q ^{-})\\
\bar{\Delta}_{\dot{{q_{ds}}}} &= D_{ds}^{-1}E_{2}'\Delta_{F_{2}} +
\begin{bmatrix} [1.8]
I_{3\times3}\\ 
\frac{\partial}{\partial q }\gamma_{ds}
\end{bmatrix}
\\
\Delta_{F_{2}} &= -(E_{2}D_{ds}^{-1}E_{2}')^{-1}E_{2}
\begin{bmatrix} [1.8]
I_{3\times3}\\ 
\frac{\partial}{\partial q }\gamma_{ds}
\end{bmatrix}
\\
E_{2}(q_{ds})&=\dfrac{\partial}{\partial q_{ds}}p_{2}(q_{ds})
\end{align*}
Here,
\begin{align*}
R=& \begin{bmatrix}
0 & 1 & 0\\
1 & 0 & 0\\
0 & 0 & 1
\end{bmatrix}
\end{align*}}

\subsection{Decoupled Dynamic Model}
Explicit expressions for the quadratic velocity forces and gravitational forces for the decoupled systems (\ref{eq:outputEQ})--(\ref{eq:actuatorEQ}) are,
\begin{align}
\tau_e=&
\begin{pmatrix}[4.3]
\begin{smallmatrix}
-l \big((4 M_H \dot{q}_1^2 r  + m \omega_{e_2}^2 r + 3 m \dot{q}_1^2 r   )\sin(\frac{\alpha}{2}) + 2 m \dot{q}_1^2 r  \sin(\beta-\frac{\alpha}{2})  \\
+ (m \dot{q}_1^2 r - 2 m \omega_{e_2} \dot{q}_1 r  )\sin(\frac{\alpha-\beta}{2})  + 2 M_T l \omega_{e_1}^2  \sin({\alpha}) \\
+( 2 m \omega_{e_2} \dot{q}_1 r - m \omega_{e_2}^2 r  - m \dot{q}_1^2 r ) \sin(\frac{\alpha+\beta}{2})\big)
\end{smallmatrix}\\
\begin{smallmatrix}
4 M_T l r \big( \sin(\frac{\alpha}{2}) +  \sin(\frac{\alpha-\beta}{2}) +  \sin(\frac{\alpha+\beta}{2})\big) \omega_{e_1}^2\\
+ r^2\big(( 8 M_H \dot{q}_1^2  + 4 M_T \dot{q}_1^2  - 2 m \omega_{e_2}^2   + 8 m \dot{q}_1^2  + 4 m \omega_{e_2} \dot{q}_1   ) \sin(\frac{\beta}{2})\\
+( 4 m \omega_{e_2} \dot{q}_1  - 2 m \omega_{e_2}^2 ) \sin({\beta})  + 2 M_T \dot{q}_1^2  \sin({\alpha})\\
+ 4 M_T \dot{q}_1^2  \sin(\alpha - \frac{\beta}{2})\big) \\
\end{smallmatrix}
\end{pmatrix} \label{eq:tau_e}
\\
\tau_z=&
\begin{smallmatrix}
M_T lr   \sin(\frac{\alpha}{2}) \omega_{e_1}^2 + \frac{r^2}{2}\big(- (m \omega_{e_2}^2   - m \dot{q}_1^2  + 2m \omega_{e_2} \dot{q}_1 )\sin(\frac{\beta}{2})\\
 +m \dot{q}_1^2  \sin({\beta}) + M_T \dot{q}_1^2  \sin({\alpha}) \big) 
\end{smallmatrix}\label{eq:tau_1}
\\
\tau_g^e=&
\begin{pmatrix}[4.2]
\begin{smallmatrix}
g l \big(2( M_H  +  M_T  +  m )\sin(q_1 + \frac{\alpha}{2} -\lambda)  \\
- m  \sin(q_1 +\frac{\alpha}{2}-\beta-\lambda) + m  \sin(q_1 -\frac{\alpha}{2} + \beta-\lambda)\\
- (2 M_H + 2 M_T +m) \sin(q_1 - \frac{\alpha}{2}-\lambda)\big)
\end{smallmatrix}\\
\begin{smallmatrix}
-2 g r \big((M_T  + M_T )\sin({\alpha} - q_1-\lambda)  + m  \sin({\beta}-q_1-\lambda)\\
 + (2 M_H + M_T + 3m )\sin(q_1 + \frac{\beta}{2}-\lambda) \\
+ (2 M_H +M_T +2m) \sin(q_1-\lambda) \\
- 2( M_H + M_T + m)\sin(q_1 - \frac{\beta}{2}-\lambda) \\
+ M_T  \sin(q_1 + \alpha -\frac{\beta}{2}-\lambda)
\big)
\end{smallmatrix}
\end{pmatrix} \label{eq:tau_ge}
\\
\tau_g^z=&
\begin{smallmatrix}
-\frac{g r}{2} \big( (2 M_H + M_T  + 2 m )\sin(q_1-\lambda) + M_T  \sin(\alpha-q_1-\lambda) \\
+ m  \sin(\beta - q_1-\lambda)\big)
\end{smallmatrix} \label{eq:tau_g1}
\end{align}

\subsection{Proof of Theorem-\ref{Theom:Main}}
\begin{proofoftheorem}
Consider the compact set $\mathcal{X}^0\triangleq \mathcal{X}_e^0\times \mathcal{Z}^0\subset \mathcal{X}$ there exists a positive definite function $W_e : TG_e \to \mathbb{R}$ with a unique minimum at the origin such that during the swing phase 
\begin{align*}
W_e(t)\leq \vartheta e^{-\kappa t},
\end{align*}
for all $x^+\in \mathcal{X}^0\triangleq \mathcal{X}_e^0\times \mathcal{Z}^0$ for some $\vartheta,\kappa>0$ while ensuring that $|\dot{q}_1(t)|\leq \xi$ for some $\xi>0$. Since $\dot{q}_1$ not zero we find that $q_1(t)$ reaches the switching surface $\mathcal{S}\subset \mathcal{Z}$ at some time $t_{k+1}^->t_k^+$.

Consider the $k^{\mathrm{th}}$ step of the robot and let $\mathcal{W}^{t_k^-}_e$ be the largest set such that
$\mathcal{W}^{t_k^-}_e\subset W_e^{-1}(\vartheta e^{-\kappa t_k^-})$. Thus we see that 
$x(t_k^-)\in\mathcal{W}^{t_k^+}_e\triangleq \mathcal{W}^{t_k^-}_e\times\{ {q_1}_{\mathrm{ref}}\}\times [-\xi,\xi]$ for any $x(t_k^+)\in \mathcal{X}^0$. Now if $x(t_{k+1}^-)=\Delta(x(t_k^-))\in \mathcal{X}^0$ then in the next step $x(t_{k+1}^-)\in\mathcal{W}^{t_{k+1}^+}_e\subset \mathcal{W}^{t_{k}^+}_e$. 
Thus we see that the trajectories of the closed-loop switched system visit $\mathcal{W}^{t_{1}^+}_e$ at the increasing sequence of time steps $t^-_0<t^-_1<\cdots<t^-_k<\cdots$. 

This set can be made arbitrarily small by picking sufficiently large gains for the PID controller and/or by making 
$\mathcal{X}^0$ sufficiently small.
Thus for any given $\delta>0$ we can pick either the gains sufficiently large or the initial condition set sufficiently small such that  for any small $\delta>0$, $\mathcal{W}^{t_{1}^+}_e\subset \mathcal{Z}_\delta $. Now it is clear that if $\mathcal{X}_0\subset \Delta(\mathcal{Z}_\delta)$ then the trajectories visit $\mathcal{Z}_\delta$ at the end of every swing phase of the robot. 
\end{proofoftheorem}

\end{appendix}
\bibliographystyle{IEEEtran}
\bibliography{Biped}

\cleardoublepage

\end{document}